\documentclass[
    aps,
    prl,
    reprint,
    superscriptaddress,
    nolongbibliography,
    nobibnotes,
    nofootinbib,
]{revtex4-2}

\usepackage[dvipsnames]{xcolor}
\definecolor{linkcolor}{rgb}{0, 0.3, 0.5}

\usepackage[
    colorlinks=true,
    linkcolor=linkcolor,
    citecolor=linkcolor,
    filecolor=linkcolor,
    urlcolor=linkcolor,
    pdfusetitle,
]{hyperref}

\usepackage{orcidlink}
\usepackage{amssymb}
\usepackage{amsmath}
\usepackage{physics}
\usepackage{mathtools}
\usepackage{cancel}

\usepackage{titlesec}
\titleformat{\section}[runin]{\itshape\bfseries}{}{0pt}{}[---]
\titlespacing{\section}{\parindent}{\parskip}{0pt}
\titleformat{\subsection}[runin]{\itshape}{}{0pt}{}[:]
\titlespacing{\subsection}{\parindent}{\parskip}{1ex}

\newcommand{\comment}[1]{}

\newcommand{\notts}{\affiliation{Nottingham Centre of Gravity \& School of Mathematical Sciences,\\University of Nottingham, University Park, Nottingham, NG7 2RD, United Kingdom}}

\begin{document}

\title{Joint population and strong-lensing inference for resolved gravitational-wave events probes the black-hole merger rate beyond the peak of star formation}

\author{Matthew Mould\,\orcidlink{0000-0001-5460-2910}}
\email{matthew.mould@nottingham.ac.uk}
\notts

\date{\today}

\begin{abstract}
Gravitational waves can be lensed by intervening potentials of various scales. Strong lensing leads to underestimated distances and overestimated masses, biasing astrophysical results if not accounted for. I present a novel analysis of the LIGO--Virgo--KAGRA catalog of binary black-hole mergers, simultaneously inferring (1) whether or not each event is strongly lensed, (2) their magnifications if so, and (3) the underlying merger population, using both parametric and nonparametric population models as well as two models for the lensing optical depth. Posterior lensing probabilities do not exceed 1\% for any event, so population constraints are consistent with those assuming nondetection of strong lensing or that lensing never occurs. This includes multiple subpopulations over black-hole mass and a component with high aligned spins. Compared to standard analyses, however, there are reductions of order 10\% in uncertainty on the redshift at which the merger rate peaks and an order of magnitude in high-redshift rate upper limits. Though modest, these are the first constraints using only resolved events at redshifts where current ground-based gravitational-wave detectors are usually insensitive, at and beyond the peak of star formation.
\end{abstract}

\maketitle

\section{Lensed gravitational waves}

The basic effect of lensing on gravitational waves (GWs) emitted by the mergers of compact objects such as black holes (BHs) is to magnify the GW strain amplitude by a factor $\sqrt{\mu}$~\cite{Schneider:1992bmb, Treu:2010uj}. As the strain amplitude is inversely proportional to luminosity distance, the apparent distance $D'$ underestimates the true distance $D=\sqrt{\mu}D'$ if lensing is not accounted for. With cosmological redshift $z(D) = z(\sqrt{\mu}D')$ in the measured detector-frame mass $m'$~\cite{LIGOScientific:2025hdt}, the source mass $m = m' / [1+z(\sqrt{\mu}D')]$ would thus be overestimated.

The LIGO--Virgo--KAGRA (LVK)~\cite{LIGOScientific:2014pky, VIRGO:2014yos, KAGRA:2020tym} catalog has over 250 detections~\cite{LIGOScientific:2026wfs} without confident lensing evidence through GWTC-4~\cite{LIGOScientific:2025slb, LIGOScientific:2025cwb}, consistent with predictions~\cite{Ng:2017yiu, Smith:2017mqu, Buscicchio:2020bdq}. A possible exception is signal distortion in GW231123\_135430 (henceforth GW231123), which has large source masses and spins~\cite{LIGOScientific:2025rsn}; waveform systematics may indicate unmodeled physics~\cite{Ray:2025rtt, Jan:2025zcm, Hu:2025lhv, Malagon:2026uev, Wang:2026yjk, Chandra:2026voe} such as lensing~\cite{Liu:2023emk, Shan:2025jpt, Chan:2025kyu, Goyal:2025eqo, Chakraborty:2025pxt, Shan:2025dcd, Barsode:2026bcs, liu2026}, but are also expected~\cite{Bini:2026kwz}.

The lack of detection does not preclude the existence of lensed GWs. Indeed, a single outlier produced by lensing would imply the existence of many others left undetected~\cite{Farah:2025ews}. In the regime of strong lensing $\mu>2$, magnified GW events above the detection threshold may have subthreshold image copies~\cite{Li:2019osa}. The accumulation of subthreshold signals increases the amplitude of the GW background~\cite{LIGOScientific:2025bgj}, which thus constrains the occurrence of lensing in resolved sources~\cite{Buscicchio:2020cij, Mukherjee:2020tvr}. Lensing rate predictions could be improved with knowledge of the source merger population~\cite{Li:2018prc}, but current constraints from GWs assume no lensing occurs~\cite{LIGOScientific:2025pvj, LIGOScientific:2026ctl}. Therefore, merger-rate constraints implied by nondetection of lensing have required inconsistent postprocessing of standard population results~\cite{, LIGOScientific:2025cwb} and lensing constraints for individual events have assumed fixed population priors~\cite{LIGOScientific:2020ufj, Liu:2020par, Lo:2021nae, Hannuksela:2025wgv, Yang:2025kfj}.

Here, this problem is solved by inferring the source population of BH mergers jointly with strong-lensing magnifications of each GW event. The 153 LVK detections through GWTC-4~\cite{LIGOScientific:2025slb} identified as BH mergers with false-alarm rates (FARs) $<1\,\mathrm{yr}^{-1}$ are included, assuming the same cosmological model as Ref.~\cite{LIGOScientific:2025hdt} throughout.

\section{Lensing model}

Following Refs.~\cite{Ng:2017yiu, Haris:2018vmn, Hannuksela:2019kle, LIGOScientific:2020ufj, Lo:2021nae}, magnifications $\mu>2$ are distributed according to $p(\mu) := p(\mu|L=1) \propto \mu^{-3}$ and the lensing probability is $P(L=1|z) = F [ D_\mathrm{c}(z) / D_\mathrm{H} ]^3$, where $D_\mathrm{c}(z)$ is the comoving distance at redshift $z$, $D_\mathrm{H}$ is the Hubble distance, and $L$ is a binary flag indicating strong lensing (1) or not (0).
To account for uncertainty on the lensing optical depth~\cite{Turner:1984ch}, both $F=0.00036$~\cite{Haris:2018vmn, LIGOScientific:2020ufj} and $F=0.0017$~\cite{Hannuksela:2019kle} are used. The probability that an event is not lensed is $P(L=0|z) = 1 - P(L=1|z)$, with magnification distribution $p(\mu|L=0) = \delta(\mu-1)$, where $\delta$ is the Dirac delta function. The joint distribution is then $p(\mu|z) = p(L=0|z) \delta(\mu-1) + p(L=1|z) p(\mu)$.

\section{Lensing population likelihood}

The GW likelihood $p(d|\theta') = p(d|\theta,\mu)$ for an event with data $d$ depends on apparent detector-frame parameters $\theta'$ (e.g., detector-frame masses and luminosity distance) \cite{Thrane:2018qnx, Talbot:2025vth, LIGOScientific:2025hdt}, which are determined by source paramters $\theta$ (e.g., source-frame masses and redshift) and magnificaton $\mu$. The source population is given in terms of a number density $\mathrm{d}N/\mathrm{d}\theta|_\Lambda$ of mergers, dependent on population-level parameters $\Lambda$. Assuming detections are realizations of an inhomogeneous Poisson process~\cite{Loredo:2004nn, Taylor:2018iat, Mandel:2018mve, Vitale:2020aaz}, the likelihood for a catalog of $N_\mathrm{obs}$ events is
\begin{align}
\mathcal{L}^l(\{\mu_i,L_i\}_{i=1}^{N_\mathrm{obs}},\Lambda)
=
e^{-N_\mathrm{exp}^l(\Lambda)} \prod_{i=1}^{N_\mathrm{obs}}
\mathcal{L}_i^l(\mu_i,L_i,\Lambda)
\, ,
\label{eq: Poisson likelihood}
\end{align}
where $L_i$ and $\mu_i$ are respectively the lensing flag and magnification for event $i$,
\begin{align}
\mathcal{L}_i^l(\mu,L,\Lambda)
=
\int \dd{\theta}
p(d_i|\theta,\mu)
p(\mu|L) p(L|\theta)
\left. \dv{N}{\theta} \right|_\Lambda
\label{eq: single event likelihood}
\end{align}
is the likelihood for event $i$ with observed data $d_i$, and
\begin{align}
N_\mathrm{exp}^l(\Lambda) =
\int \dd{d} \dd{\theta} \dd{\mu}
S(d) p(d|\theta,\mu) 
p(\mu|\theta)
\left. \dv{N}{\theta} \right|_\Lambda
\label{eq: Nexp}
\end{align}
is the expected number of events, with a threshold $S(d)$ on the data selecting $\mathrm{FAR}(d)<1\,\mathrm{yr}^{-1}$. Due to the nondetection of multiple images in targeted searches~\cite{LIGOScientific:2025cwb}, here we are continuing the assumption that any event that is strongly lensed has subthreshold signal copies, so events are otherwise independent of each other. This likelihood can thus been seen as the simplest generalization of the strong-lensing analysis in, e.g., Ref.~\cite{LIGOScientific:2020ufj}.

The typical approach to compute Eqs.~\eqref{eq: single event likelihood} and \eqref{eq: Nexp} is Monte Carlo integration, reusing parameter-estimation (PE) \cite{Ashton:2025xba} and software-injection \cite{Tiwari:2017ndi} samples from public LVK data products \cite{LIGOScientific:2025snk}. Throughout, waveforms for PE results are selected as in Ref.~\cite{Mould:2026nle}, the survey sensitivity is estimated following Ref.~\cite{Essick:2025zed}, and uncertainty in the estimated likelihood due to Monte Carlo approximations~\cite{Tiwari:2017ndi, Farr:2019rap, Essick:2022ojx, Talbot:2023pex, Heinzel:2025ogf} is limited following Ref.~\cite{LIGOScientific:2025pvj}. However, these data products assume $\mu=1$ for all signals, so sources must be defined in terms of apparent detector-frame parameters $\theta'$, not the source parameters $\theta$.

Bayes' theorem gives $p(d|\theta') \propto q(\theta'|d) / q(\theta')$, where $q(\theta'|d)$ is the single-event posterior for detector-frame parameters inferred under the standard PE prior $q(\theta')$~\cite{LIGOScientific:2025slb}. Substituting and changing variables in Eq.~\eqref{eq: single event likelihood} gives
\begin{align}
\mathcal{L}_i^l(\mu,L,\Lambda)
\! = \!\!
\int \! \dd{\theta'} \!
\left| \pdv{\theta}{\theta'} \right|_\mu \!\!
\frac{q(\theta'|d_i)}{q(\theta')}
p(\mu|L) p(L|\theta) \!
\left. \dv{N}{\theta} \right|_\Lambda
\end{align}
where here and in the following it is left implicit that $\theta$ depends on $\theta'$ and $\mu$. This integral can now be computed with a Monte Carlo approximation over samples from $q(\theta'|d_i)$ for each event.

For Eq.~\eqref{eq: Nexp}, simulated signals are drawn from a fixed reference distribution $r(\theta)$, which implies a fixed apparent detector-frame distribution $r(\theta') = r(\theta) | \mathrm{d}\theta/\mathrm{d}\theta' |_{\mu=1}$. The model for $p(\mu|\theta) = p(\mu|z)$ above implies that $N_\mathrm{exp}^l(\Lambda) = N_\mathrm{exp}^{l,0}(\Lambda) + N_\mathrm{exp}^{l,1}(\Lambda)$, where after substitution and changing variables the expected numbers of nonlensed and lensed events are respectively
\begin{align}
N_\mathrm{exp}^{l,0}(\Lambda) & =
\int \dd{d} \dd{\theta'}
S(d) p(d|\theta')
r(\theta') w_0(\theta',\Lambda)
\, , \\
N_\mathrm{exp}^{l,1}(\Lambda) & =
\int \dd{d} \dd{\theta'} \dd{\mu}
S(d) p(d|\theta')
r(\theta') p(\mu) w_1(\mu,\theta',\Lambda)
\, . \nonumber
\end{align}
These integrals can now be computed with Monte Carlo approximations over samples from the detected software injections---$p(d|\theta') r(\theta') \delta(\mu-1)$ for the first and $p(d|\theta') r(\theta') p(\mu)$ for the second with randomly assigned magnifications from $p(\mu)$---using weights
\begin{align}
w_0(\theta',\Lambda)
& =
\frac{p(L=0|\theta)}{r(\theta')}
\left. \dv{N}{\theta} \right|_\Lambda
\left|\pdv{\theta}{\theta'}\right|_{\mu=1}
, \\
w_1(\mu,\theta',\Lambda)
& =
\frac{P(L=1|\theta)}{r(\theta')}
\left. \dv{N}{\theta} \right|_\Lambda
\left|\pdv{\theta}{\theta'}\right|_\mu
.
\end{align}

In all the above, the magnification-dependent Jacobian for the conversion between intrinsic source parameters $\theta$ and apparent detector-frame parameters $\theta'$ is
\begin{align}
\left|\pdv{\theta'}{\theta}\right|_\mu
& =
\left|\pdv{(m_1',m_2',D')}{(m_1,m_2,z)}\right|_\mu
=
\frac{(1+z)^2}{\sqrt{\mu}} \dv{D}{z}
\, ,
\end{align}
where $m_1 \geq m_2$ ($m_1' \geq m_2'$) are the source (apparent detector-frame) primary and secondary masses.

\section{Lensing nondetection}

Given that targeted strong lensing searches find no significant detections \cite{ LIGOScientific:2025cwb}, a reasonable assumption is instead that $L_i=0$ and thus $\mu_i=1$ for each event $i=1,...,N_\mathrm{obs}$. The likelihood in this case is $\mathcal{L}^l ( \{\mu_i=1,L_i=0\}_{i=1}^{N_\mathrm{obs}} , \Lambda)$.

\section{No lensing}

Results are also compared to those under the standard assumption that lensing never happens~\cite{LIGOScientific:2018jsj, LIGOScientific:2020kqk, KAGRA:2021duu, LIGOScientific:2025pvj, LIGOScientific:2026ctl}. The population likelihood is
\begin{align}
\mathcal{L}^n(\Lambda) \propto
e^{-N_\mathrm{exp}^n(\Lambda)}
\prod_{i=1}^{N_\mathrm{obs}}
\mathcal{L}_i^n(\Lambda)
\, .
\end{align}
The single-event likelihoods are
\begin{align}
\mathcal{L}_i^n(\Lambda) \propto
\int \dd{\theta'}
\left| \pdv{\theta}{\theta'} \right|_{\mu=1}
\frac{q(\theta'|d_i)}{q(\theta')}
\left. \dv{N}{\theta} \right|_\Lambda
.
\end{align}
The expected number of detections is
\begin{align}
N_\mathrm{exp}^n(\Lambda) &=
\int \dd{d} \dd{\theta'}
S(d) p(d|\theta') r(\theta')
w(\theta',\Lambda)
\, , \\
w(\theta',\Lambda) &=
\frac{1}{r(\theta')}
\left. \dv{N}{\theta} \right|_\Lambda
\left|\pdv{\theta}{\theta'}\right|_{\mu=1}
.
\end{align}
These integrals also admit Monte Carlo approximations.

Note that the no-lensing likelihood is distinct from the nondetection likelihood. The population posterior inferred using the nondetection likelihood is also not the same as filtering out populations that overpredict lensed detections from the no-lensing posterior, an approximation~\cite{LIGOScientific:2025cwb} that is only valid if $P(L=0|z) \approx 1$.

\section{Population models}

One parametric model and one nonparametric model is used for the density $\mathrm{d}N/\mathrm{d}\theta = p(m_1) p(m_2|m_1) p(\chi_\mathrm{eff}) \mathrm{d}N/\mathrm{d}z$ of mergers over component masses $m_{1,2}$, effective spin $\chi_\mathrm{eff}$~\cite{Racine:2008qv, LIGOScientific:2025hdt}, and redshift $z$.

\subsection{Parametric}

For BH masses, the \textsc{Broken Power Law + 2 Peaks} model is used~\cite{LIGOScientific:2025pvj} but modified as in Ref.~\cite{Mould:2026sww}, with the priors in their Table~IV except for a uniform prior on the maximum BH mass between $100M_\odot$ and $200M_\odot$.

The spin population distribution $p(\chi_\mathrm{eff})$ is assumed to be Gaussian over $\chi_\mathrm{eff} \in [-1,1]$, with uniform priors on the mode $\in[-1,1]$ and scale parameter $\in[0,1]$.

The BH merger rate per unit comoving volume $V_\mathrm{c}$ and source-frame time $t_\mathrm{s}$ follows a Madau Dickinson type evolution over redshift~\cite{Madau:2014bja, Madau:2016jbv, Fishbach:2018edt, Callister:2020arv}:
\begin{align}
\frac{\mathrm{d}N}{\mathrm{d}V_\mathrm{c}\,\mathrm{d}t_\mathrm{s}}
=
\mathcal{R}_0
\left[ 1 + (1+z_\mathrm{p})^{-\gamma-\kappa} \right]
\frac
{ (1 + z)^\gamma }
{ 1 + \left( \frac{1+z}{1+z_\mathrm{p}} \right)^{\gamma+\kappa} }
\, ,
\end{align}
where $\gamma$ is the slope at low redshift, $\kappa$ the slope at high redshift, and $z_\mathrm{p}$ the redshift at which the rate turns over. This is converted to a number density over redshift according to
\begin{align}
\dv{N}{z} =
\frac{\mathrm{d}N}{\mathrm{d}V_\mathrm{c}\,\mathrm{d}t_\mathrm{s}}
\dv{V_\mathrm{c}}{z}
\frac{T}{1+z}
\, ,
\end{align}
where $T$ is the observing time of the detectors. Uniform priors are taken for $z_\mathrm{p}\in[0,5]$, $\gamma\in[-10,10]$, and $\kappa\in[0,10]$, while the standard improper log-uniform prior is used for the merger rate $\mathcal{R}_0$ at $z=0$~\cite{Fishbach:2018edt, LIGOScientific:2025pvj}.

\subsection{Nonparametric}

To account for unknown population features not captured by these simple functional forms, a more flexible model is also considered for $\mathrm{d}N/\mathrm{d}V_\mathrm{c}\,\mathrm{d}t_\mathrm{s}$ as a function of $z$ and the one-dimensional population densities of $p(\ln{m_1})$, $p(\ln{m_2}|\ln{m_1})$, and $p(\chi_\mathrm{eff})$. These are all taken to be piecewise linear functions at the same time, with 100 nodes placed linearly over $z\in[0,5]$, $\ln{m_{1,2}}\in[\ln{3},\ln{200}]$, and $\chi_\mathrm{eff}\in[-1, 1]$.

The intrinsic conditional autoregressive (ICAR) model~\cite{0abfb364-d2b0-3c36-bea9-11d72a38fbef, d0d71b5d-0bf6-3ac1-abe8-ba4e368b1424} is used as a minimal smoothing prior for each function, as in the \textsc{PixelPop} model~\cite{Heinzel:2024jlc, Heinzel:2024hva, Alvarez-Lopez:2025ltt, LIGOScientific:2026ctl}. This is an improper zero-mean multidimensional Gaussian prior on the differences between function values at neighboring nodes. For each function, there is a free parameter $\sigma$ that controls the variance of these values. The ICAR prior can be marginalized in closed form over a gamma prior on $\sigma$. In particular, an improper log-uniform prior is used, which is the Jeffrey's prior~\cite{Jeffreys:1939xee, 10.1098/rspa.1946.0056, gelman1995bayesian}.

\section{Results}

In summary, there are ten population analyses: lensing and nondetection with $F=0.00036$ and $F=0.0017$, and standard no-lensing, with all cases repeated using parametric and nonparametric models. Inference is performed using \textsc{gwax}~\cite{gwax}, with \textsc{numpyro}~\cite{numpyro} for sampling in \textsc{jax}~\cite{jax}.

\textbf{There is no evidence for strongly lensed events.}
Across the four lensing analyses, no event exceeds a posterior lensing probability of 1\%. For the parametric population models, the event with highest probability is GW230704\_212616, at 0.3\% for $F=0.00036$ and 0.7\% for $F=0.0017$. The two nonparametric analyses find this event to have the second highest probability $\lesssim0.5\%$. Under the strong-lensing assumption, this event supports magnifications $\lesssim6$. The initial PE posterior for this event is that with the highest support for high luminosity distances in the catalog~\cite{LIGOScientific:2025slb}, and without support for large magnifications, this translates to the highest redshift and thus the highest probability of lensing.

\begin{figure}
\centering
\includegraphics[width=1\columnwidth]{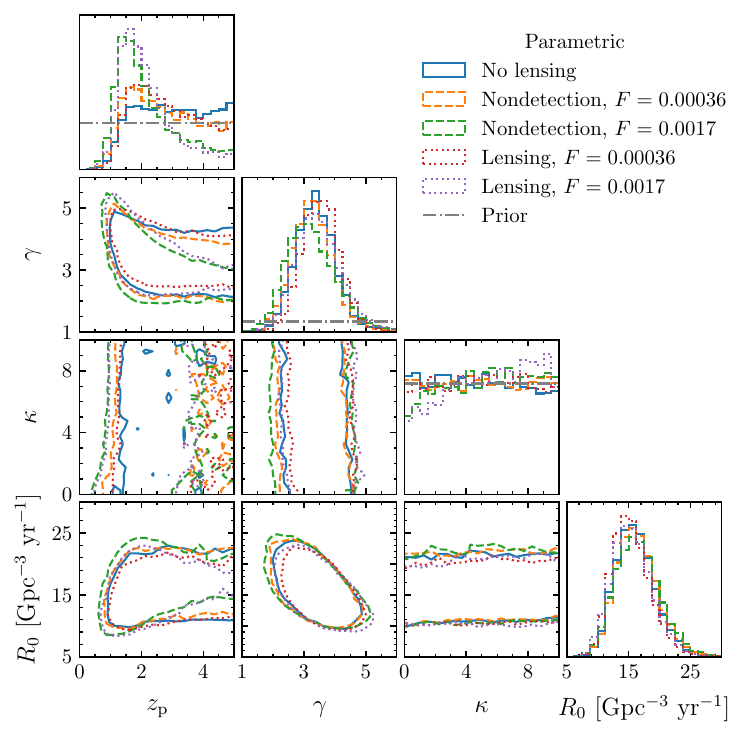}
\caption{Posterior distributions for the population parameters governing the evolution of the merger rate over redshift. Diagonal panels show one-dimensional marginals and other panels show 90\% credible regions for two-dimensional marginals.}
\label{fig: corner}
\end{figure}

\textbf{Nondetection of strong lensing constrains the BH merger rate at high redshifts.} Posterior distributions for the parameters of the parametric merger-rate model are shown in Fig.~\ref{fig: corner}. The posterior medians and 90\% equal-tailed intervals for the redshift where the rate peaks are $z_\mathrm{p} = 2.0_{-0.9}^{+2.2}$ and $z_\mathrm{p} = 2.7_{-1.3}^{+2.0}$ for the $F=0.0017$ and $F=0.00036$ strong-lensing population analyses, respectively. Though the posteriors do not rule out any additional redshifts, they give modest 12\% and 4\% reductions in uncertainty in terms of 90\% intervals, or 19\% and 6\% reductions in terms of $1\sigma$ uncertainties, compared to $z_\mathrm{p} = 3.0_{-1.7}^{+1.8}$ from standard population analyses. Whereas standard results do not constrain $z_\mathrm{p}$ except to rule out low redshifts, these constraints feature a prominent posterior peak.

As there is no evidence for strong lensing, these population constraints are equivalent to those assuming nondetection of strong lensing (compare dashed and dotted distributions in Fig.~\ref{fig: corner}). Other population parameters are essentially identical also to those inferred from the standard no-lensing analysis, which finds a local merger rate $R_0 = 15.9_{-4.5}^{+6.0} \, \mathrm{Gpc}^{-3} \, \mathrm{yr}^{-1}$, a slope below $z_\mathrm{p}$ of $\gamma = 3.4_{-1.0}^{+1.1}$, and an unconstrained slope $\kappa$ above $z_\mathrm{p}$.

\begin{figure*}
\centering
\includegraphics[width=1\textwidth]{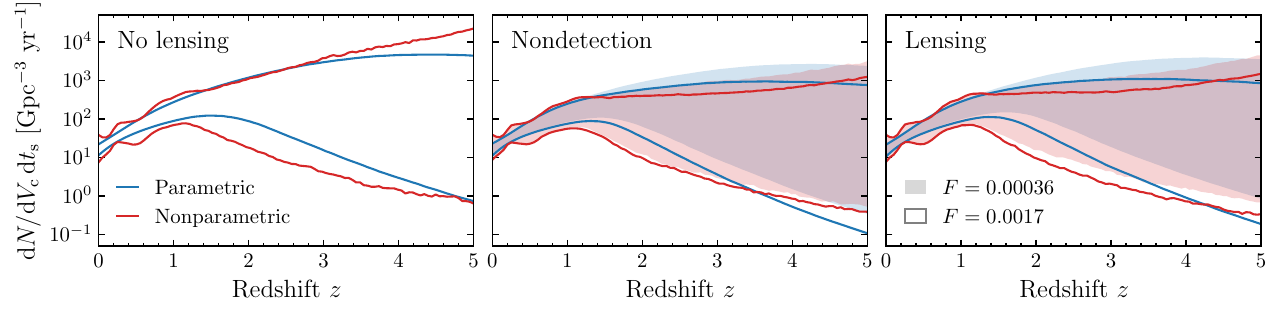}
\caption{Inferred evolution of the merger rate per unit comoving volume and source-frame time as a function of redshift. Results using the parametric and nonparametric models are in blue and red, respectively. Results are from standard population analyses (left), assuming nondetection of strong lensing (middle), and inferring strong lensing for all events (right). For the latter two cases, two variations of the lensing optical depth model with lower (shaded) and higher (solid) prior lensing probabilities are given. Regions in all panels enclose 90\% equal-tailed credible intervals.}
\label{fig: ppd-z}
\end{figure*}

\begin{figure*}
\centering
\includegraphics[width=1\textwidth]{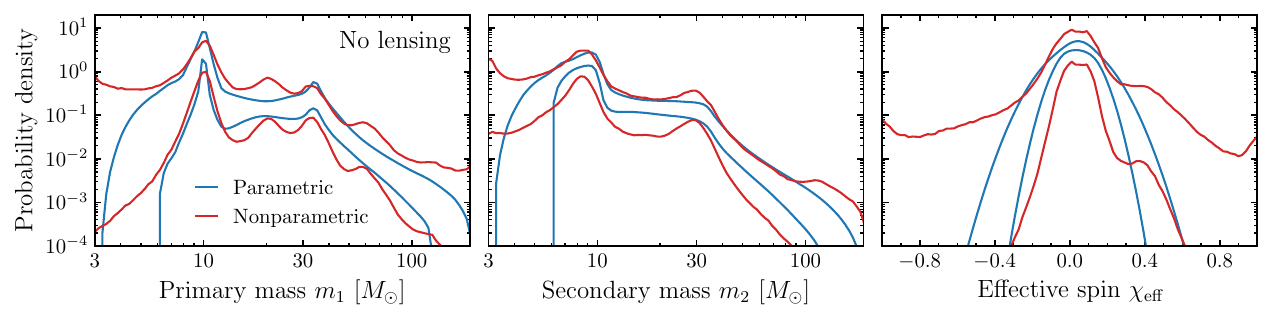}
\caption{Posterior population-level distributions of logarithmic primary BH mass (left), logarithmic secondary BH mass (middle), and effective spin (right) inferred from standard GW population analyses assuming no lensing. Results for the nondetection and lensing analyses are almost identical to these. The parametric and nonparametric models are shown in blue and red, respectively. Enclosed regions are 90\% equal-tailed credible intervals as a function of the source parameter in each panel.}
\label{fig: ppd-others}
\end{figure*}

Figure~\ref{fig: ppd-z} compares the parametric and nonparametric merger-rate constraints as a function of redshift for the no-lensing, nondetection, and lensing population analyses. The latter two again produce similar results. Nonparametric uncertainties are typically larger than parametric ones, reflecting a bias--variance tradeoff with increased model flexibility, but suggest possible nonmonotonic evolution at redshifts $z<1$~\cite{Edelman:2022ydv, Godfrey:2023oxb, Callister:2023tgi, Heinzel:2024hva, LIGOScientific:2026ctl}. High-redshift constraints on the rate span several orders of magnitude in all cases. However, the nondetection and lensing analyses produce more stringent upper bounds. For example, the no-lensing nonparametric (parametric) analysis finds rates $\lesssim 2.2\times10^4$\,($4.4\times10^3$)\,Gpc$^{-3}$\,yr$^{-1}$ at $z=5$ at 95\% credibility, compared to the most stringent limits $\lesssim1.2\times10^3$\,($7.6\times10^2$)\,Gpc$^{-3}$\,yr$^{-1}$ coming from the $F=0.0017$ nondetection analysis. Over redshifts $1<z<4$, the nonparametric model in fact provides more stringent upper 95\% credible levels than the parametric model.

\textbf{The inferred BH merger population is robust to strong lensing.} This is confirmed by jointly inferring strong-lensing magnifications and the underlying BH merger population for the first time. There are no differences in the population-level distributions of binary BH masses and spins inferred from the no-lensing, nondetection, and lensing analyses. All three cases are therefore illustrated with the no-lensing results in Fig.~\ref{fig: ppd-others}.

There are four potential overdensities at $\sim10M_\odot$, $20M_\odot$, $30M_\odot$, and $60M_\odot$ on top of the primary BH mass spectrum that may represent distinct subpopulations~\cite{Tiwari:2026bvx, Edelman:2022ydv, Godfrey:2023oxb, Toubiana:2023egi, Callister:2023tgi, Guttman:2025jkv, Banagiri:2025dmy, Ray:2026uur, MaganaHernandez:2024qkz, Wang:2025nhf, Pierra:2026ffj, Bertheas:2026odj, Gennari:2026dfy, Cheng:2026bpc, Galaudage:2026opk}. The secondary-mass distribution features prominent peaks only at $\sim10M_\odot$ and $30M_\odot$~\cite{Farah:2023swu, Heinzel:2024hva, Mould:2026sww, Li:2026amt}, suggestive of the additional primary-mass peaks being imprints of repeated BH mergers~\cite{Gerosa:2021mno, Kimball:2020qyd, Mould:2022ccw, Li:2023yyt, Tong:2025wpz, Antonini:2025ilj, Tong:2025xir, Plunkett:2026pxt, Farah:2026jlc, Vijaykumar:2026zjy, Ginat:2026awh}. However, one should be careful not to overinterpret population features that may arise due to the finite catalog of observations~\cite{Corelli:2026thw}; e.g., the upper credible levels for masses $\gtrsim100M_\odot$ are driven by GW231123~\cite{Mould:2026sww}.

There is also a potential increase in the merger rate for sources with $\chi_\mathrm{eff}\sim0.4$, requiring binaries with large BH spins preferentially aligned to the orbital angular momentum~\cite{Wang:2021clu, Li:2025rhu, Li:2025iux, Alvarez-Lopez:2026ymo, Flanagan:2026ayy, Padhyegurjar:2026scg, rinaldi2026}. That the inferred spin distribution is unchanged with the inclusion of strong lensing is expected, as the effect of strong lensing included here has no impact of BH spin. However, this need not be the case at the population-level if masses and spins are correlated~\cite{Callister:2021fpo, Biscoveanu:2022qac, Heinzel:2023hlb, Pierra:2024fbl, Antonini:2024het, Berti:2025usa, Stegmann:2025zkb, Wolfe:2026meb, Biscoveanu:2026ikx} and lensing appreciably impacts the mass distribution, similar to cosmology~\cite{Tong:2025xvd}.

\section{Discussion}

Though strongly lensed GWs could produce systematically overestimated source masses and thus biased population estimates, the new joint population--lensing analyses presented here show that there is currently no evidence for this, in contrast to previous claims~\cite{Broadhurst:2018saj, Diego:2021fyd}. These results are therefore compatible with post-hoc consistency checks for lensing as an origin of high-mass mergers~\cite{Harshe:2026wcj} and the nondetection of strong lensing in targeted searches~\cite{LIGOScientific:2025cwb}. High-mass mergers like the source of GW231123 are not found to have high lensing probabilities and magnifications, implying strong lensing is not necessary to make them consistent with the astrophysical population. Indeed, GW231123 was not found to be an outlier~\cite{LIGOScientific:2025rsn, LIGOScientific:2025pvj, Mandel:2025qnh, Tenorio:2026dcc} and its source properties are reproduced by many different astrophysical formation models~\cite{Stegmann:2025cja, Tanikawa:2025fxw, Bartos:2025pkv, Croon:2025gol, Delfavero:2025lup, Gottlieb:2025ugy, Popa:2025dpz, Kiroglu:2025vqy, Paiella:2025qld, Passenger:2025acb, Roupas:2026anb, Angeloni:2026nmy}. 

The simple piecewise lensing model compatible with previous works could be improved with more detailed models, in particular by using a magnification distribution that depends continuously on the merger redshift~\cite{Dai:2016igl}. Furthermore, the possibility that signals in the GW catalog are copies of each other was not accounted for here. Including multiple-image inference~\cite{Lo:2021nae} jointly with population inference would be an interesting extension to this first step. The constraints here could additionally be combined with upper limits from the stochastic background that account for lensing~\cite{Buscicchio:2020cij} in a joint inference set up, rather than checking compatibility after the fact~\cite{Harshe:2026wcj}.

High-redshift constraints on the BH merger rate are usually only accessible by considering subthreshold GW signals, e.g., from upper limits on the stochastic GW background~\cite{Buscicchio:2020cij, LIGOScientific:2025bgj, Callister:2020arv, Sah:2025agw}. Unlike previous population-level constraints, the analyses here demonstrate that joint strong-lensing and population inference offers comparable constraints at and beyond the peak of star formation $z\sim2$--3. These are the first analyses that use only resolved GW events in the LVK catalog to constrain the BH merger rate at such redshifts, where standard analyses are usually not sensitive. Inference of the peak and high-redshift rate will have implications for the evolutionary histories of binary mergers, in particular via their time delays~\cite{Fishbach:2023pqs, Schiebelbein-Zwack:2026sqj, Padhyegurjar:2026slt, Mukherjee:2021qam}, and thus on the formation channels compatible with such rates~\cite{Boco:2026qui}. This method thus provides a new probe of GW sources.

\vspace{\baselineskip}
\section{Acknowledgments}

I thank Amanda Farah, Stephen Green, Otto Hannuksela, and Anarya Ray for discussions.
I am supported by a Research Fellowship from the Royal Commission for the Exhibition of 1851
and acknowledge computational resources provided by
the University of Nottingham Centre of Gravity and Ada HPC service,
and the LIGO Laboratory supported by National Science Foundation Grants PHY-0757058 and PHY-0823459.
This research has made use of data or software obtained from the Gravitational Wave Open Science Center (gwosc.org), a service of the LIGO Scientific Collaboration, the Virgo Collaboration, and KAGRA,
and is based upon work supported by NSF's LIGO Laboratory, which is a major facility fully funded by the National Science Foundation.

\bibliography{draft}

\end{document}